\documentclass[twocolumn,floats,showpacs,prl,amssymb,floatfix,nofootinbib,balancelastpage,superscriptaddress,amsmath]{revtex4}
\usepackage{epsfig}

\newcommand{\<}[1]{\langle {#1} \rangle}

\def\sss{\scriptscriptstyle}

                              \newlength{\strikewidth}
                              \newlength{\strikelength}
\begin{document}

\title{What mass are the smallest protohalos?}

\author{Stefano Profumo}
\email{profumo@caltech.edu}
\affiliation{California Institute of Technology, Mail Code
130-33, Pasadena, CA 91125}
\author{Kris Sigurdson}\thanks{Hubble Fellow}
\email{krs@ias.edu}
\affiliation{Institute for Advanced Study, Einstein Drive, Princeton, NJ 08540}
\author{Marc Kamionkowski}
\email{kamion@tapir.caltech.edu}
\affiliation{California Institute of Technology, Mail Code
130-33, Pasadena, CA 91125}


\begin{abstract}
\noindent 
We calculate the kinetic-decoupling temperature for
weakly interacting massive particles (WIMPs) in supersymmetric (SUSY)
and universal-extra-dimension (UED) models that can
account for the cold-dark-matter abundance determined from
cosmic microwave background measurements.
Depending on the parameters of the particle-physics model,
a wide variety of decoupling temperatures is possible, ranging
from several MeV to a few  GeV. These decoupling temperatures
imply a range of masses for the smallest protohalos much larger than previously thought --- ranging from
$10^{-6}~{\rm M}_{\oplus}$ to  $10^{2}~{\rm M}_{\oplus}$.
We expect the range of protohalos masses
derived here to be characteristic of most particle-physics
models that can thermally accommodate the required relic
abundance of WIMP dark matter, even beyond SUSY and UED.

\end{abstract}




\maketitle

The physical nature of dark matter remains one of the major
unsolved problems in theoretical physics and cosmology.  
One of the leading candidates for dark matter is a
weakly interacting massive particle (WIMP)
\cite{WIMPReviews}.  In the simplest models, WIMPs (which we
denote by $X$) carry a conserved quantum number that renders
them stable.  When the
temperature in the early Universe drops below $m_{X}$, the WIMP abundance
creeps down the Boltzmann tail until pairs of WIMPs
can no longer find each other within a Hubble time, and
the comoving number density of WIMPs becomes constant.  Up to
factors of a few, this freeze-out of the annihilation channel
happens at a temperature $T_{\rm fo} \sim m_{X}/20$ and leads to a
relic abundance of dark matter of $\Omega_{X} h^2 \simeq (3
\times 10^{-27}~{\rm cm}^3{\rm s}^{-1})/\<{\sigma_a v}$, where
$\<{\sigma_a v}$ is the thermally averaged cross section (times
relative velocity) for annihilation of $X$ pairs into lighter
particles.  In typical models, $m_{X} \sim 100-1000$~GeV, and
$T_{fo} \sim 5-50$~GeV.

While freeze-out signals the departure of WIMPs from
{\em chemical} equilibrium, it does \emph{not} signal the end of WIMP
interactions.  Elastic and inelastic scattering processes of the
form $X f \rightarrow X f$ or $X f \rightarrow X^{\prime}
f^\prime$ keep the dark matter in \emph{kinetic}
equilibrium until later times (lower temperatures)
\cite{Boehm,Chen:2001jz,GreenWIMPy}. Here $f$ and $f^\prime$ are
SM particles in the thermal bath (leptons, quarks, gauge bosons) and
$X^{\prime}$ is an unstable particle that carries the same
conserved quantum number as $X$.  The temperature $T_{\rm kd}$ of {\em
kinetic} decoupling sets the
distance scale at which linear density perturbations in the
dark-matter distribution get washed out---the small-scale 
cutoff in the matter power spectrum.  In turn, this small-scale
cutoff sets the mass $M_c$ of the smallest protohalos that
form when these very small-scales go nonlinear at a redshift $z \sim 70$.
There may be implications of this small-scale cutoff for direct
\cite{Diemand:2005vz} and indirect \cite{Ando:2005xg} detection.

Some early work assumed that the cross sections for WIMPs to
scatter from light particles (e.g., photons and neutrinos) would
be energy independent, leading to suppression of power out to
fairly large (e.g., galactic) scales.  However, in
supersymmetric models, at least, the relevant elastic-scattering
cross sections drop precipitously with temperature, resulting in
much higher $T_{\rm kd}$ and much smaller suppression
scales \cite{Chen:2001jz}.  If the
annihilation cross section of WIMPs into light fermions goes as
$\sigma_{a} \simeq g_a^4/m_{X}^2$, then one expects the
scattering cross section to be $\sigma_{s} \simeq g_{s}^4
E^2/m_X^4$,
where $E$ is the energy of the scattering light particles, and
$g_a \sim g_s$ up to factors of order unity.  This estimate has
been used to derive $T_{\rm kd}$ and infer
that the minimum protohalo mass is $M_c \sim {\rm M}_{\oplus}$
\cite{GreenWIMPy,Loeb:2005pm,Diemand:2005vz}.  However, to date,
no detailed calculation of $T_{\rm kd}$ and $M_c$
in supersymmetric or other models consistent with experimental
and cosmological data have been performed.

In this \emph{Letter}, we calculate the kinetic-decoupling
temperature $T_{\rm kd}$ of WIMP dark matter in models
that account
for the correct cold-dark-matter density while remaining
consistent with laboratory constraints.
We consider models within the minimal
supersymmetric extension of the standard model (MSSM) and
models with universal extra dimensions (UED).  Instead of
relying on heuristic arguments or toy models, we use the
detailed scattering cross sections of WIMPs, including
resonances and threshold effects, both for the
WIMP relic abundances and for $T_{\rm kd}$. The main result of
our analysis is that $T_{\rm kd}$ may range all the way from
tens of MeV to several GeV.  These $T_{\rm kd}$ imply a range
$M_c\sim10^{-6}~{\rm M}_{\oplus}$ to $M_c\sim10^{2}~{\rm
M}_{\oplus}$, where we use the estimate \cite{Loeb:2005pm}
\begin{equation}
     M_c \simeq 33.3\left(T_{\rm kd}/10\ {\rm
     MeV}\right)^{-3}~{\rm M}_{\oplus},
\end{equation}
which accounts for both the acoustic oscillations imprinted on
the power spectrum by the coupling between the dark
matter and the relativistic particles in the primordial plasma
prior to kinetic decoupling and the cutoff due to free-streaming
of dark matter after kinetic decoupling.  Although we focus
on particular WIMP scenarios, we expect the range of $M_c$
derived here will be characteristic of most
particle-physics models that can accommodate the required relic
abundance of thermal WIMP dark matter.

We define $T_{\rm kd}$ from
$\tau_r(T_{\rm kd})=H^{-1}(T_{\rm kd})$
\cite{GreenWIMPy}, where $H(T)$ is the Hubble expansion rate,
and the relaxation time $\tau_r$ is
\begin{equation}
    \tau_r^{-1}\equiv \sum_l
    n_{l}(T,m_l)\sigma_{lX}(T)(T/m_X). 
\end{equation}
Here, $n_{l}(T,m_l)\sim T^3$ is the equilibrium number density
of the relativistic particle species $l$ (the true mass
dependence can be crucial here for some of the species under
consideration such as the $\mu$ and $\tau$ leptons),
$\sigma_{lX}(T)$ is the thermally averaged scattering cross
section of the WIMP $X$ off $l$'s, and the factor $(m_X/T)^{-1}$
counts the number of scatters needed to
keep the WIMPs in kinetic equilibrium. Here, we consider $l \in
\{ \nu_{e,\mu,\tau},\ e^{\pm},\
\mu^{\pm},\  \tau^{\pm} \}$ and neglect the scattering off
light quarks.  This is well justified for temperatures $T_{\rm kd}
\ll m_\pi < \Lambda_{QCD}$, as the scattering of WIMPs off mesons
and hadrons is suppressed with respect to their scattering off
light leptons by the relative abundance of the species in the
thermal bath.  In fact, in some cases, we find
$T_{\rm kd} \gtrsim m_\pi$, and
strictly speaking in these cases a detailed model for the confinement
mechanism should be included. Here, we neglect
these effects and give what can be regarded, in this regime, as
an upper limit to $T_{\rm kd}$ (taking into account the scattering
off strongly-interacting particles would in fact {\em decrease}
$T_{\rm kd}$).
\begin{figure}
\centerline{\epsfig{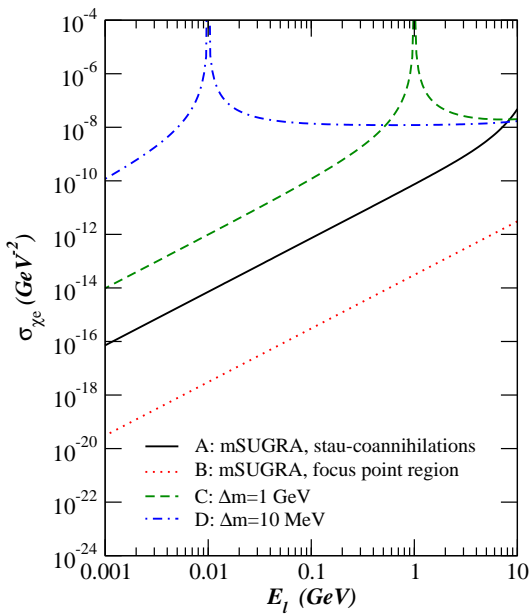}}
\caption{Neutralino-electron scattering cross section as a
    function of the electron energy $E_l$ for the four benchmark
    models {\bf A}-{\bf D} discussed in the text.}
\label{fig:sigmachie}
\end{figure}

In the case of supersymmetric models, the scattering of
neutralinos ($\chi$'s) off leptons proceeds through sfermion
and gauge-boson exchange. The relevant cross sections have been
computed in Ref.~\cite{Chen:2001jz} for the case of
neutralino-neutrino scattering. We extend here the results of
Ref.~\cite{Chen:2001jz} to include charged-lepton scattering,
where further diagrams (involving both right- and left-handed
charged-slepton exchange) as well as novel interfering
amplitudes appear. The scattering cross section $\sigma_{\chi
l}$ goes as $E_l^2$ \cite{Chen:2001jz},
modulo resonant channels where the exchanged slepton mass is
quasi-degenerate with the neutralino mass. In this latter case,
the slepton width has to properly be taken into account in the
computation of $\sigma_{\chi l}(T)$.  

We show our results for the neutralino-electron scattering cross
section as a function of energy in Fig.~\ref{fig:sigmachie},
where we pick supersymmetric ``benchmark'' models in the
context of the minimal supergravity (mSUGRA) paradigm
\cite{msugra}. We set for all models (with the usual notation)
$m_{1/2}=500$ GeV, $A_0=0$, $\tan\beta=10$, $\mu>0$, and
$m_t=172.7$ GeV; in all cases, the neutralino mass is around 200
GeV. Model {\bf A} features $m_0=100$ GeV, and lies in the {\em
coannihilation region}, where scalar superparticles are light,
and the next-to-lightest supersymmetric particle (NLSP) is a
$\tau$ slepton. The latter is, here, quasi-degenerate with the
lightest neutralino, and coannihilation among the two
species brings the neutralino relic abundance into accord
with the dark-matter abundance. Model {\bf B}
belongs, instead, to the {\em focus-point region}, where large
scalar masses (here, $m_0=2770$ GeV) at the grand-unification
scale drive the higgsino mass parameter $\mu$ to low values at
the weak scale through renormalization-group evolution and
radiative electroweak-symmetry breaking. A low value of $\mu$
implies a mixed higgsino-bino dark-matter particle, which again
can produce a thermal relic abundance in the cosmological
density range. Heavy sfermions imply that scattering off light
fermions proceeds through $Z^0$ exchange, and the resulting
$\sigma_{\chi l}$ is suppressed with respect to the
light-sfermion case (model {\bf A}) by almost four orders of 
magnitude.

We also examine models that exhibit the effects of sfermion
resonances in neutralino-lepton scattering. We modify model
{\bf A}, lowering the soft-supersymmetry-breaking left-handed
slepton masses of the first two generations, in order to get
$\Delta m_{\widetilde \nu_{e,\mu}}\equiv m_{\widetilde
\nu_{e,\mu}}-m_\chi\simeq\Delta m_{\widetilde e_1,\widetilde
\mu_1}=1$ GeV (model {\bf C}) and 0.01 GeV (model {\bf
D}). These models can be motivated in the context of
extensions of MSUGRA with non-universal scalar masses (see, e.g.,
Ref. \cite{nonuniv}).  At sufficiently small
temperatures, $\sigma_{\chi l}\propto E_l^2$ is recovered,
but for $T\gtrsim\Delta m_{\widetilde \nu}$, $\sigma_{\chi
l}\simeq$constant, and is simply set by the neutralino mass and
by the relevant neutralino-lepton-slepton couplings ($\sigma_{\chi
l}\propto |g_{\chi \widetilde l l}|^4/m_{\chi}^2$).

\begin{figure}
\centerline{\epsfig{file=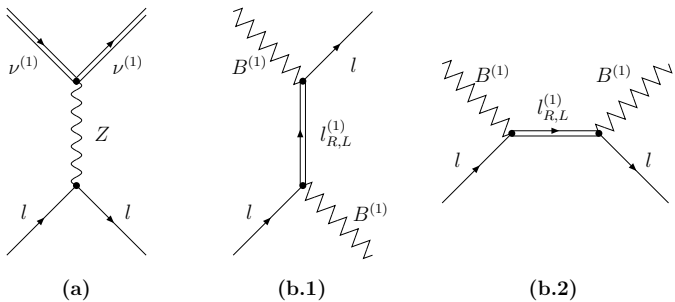,width=3.5in,angle=0}}
\caption{Feynman diagrams contributing to the scattering of
     $\nu^{(1)}$ (a) and $B^{(1)}$ (b.1 and b.2) off leptons.} 
\label{fig:feyn}
\end{figure}

Another class of WIMP models that has recently received
considerable attention is that arising in the context of
universal extra dimensions (UED) \cite{ued}. In minimal setups,
UED features a stable lightest Kaluza-Klein particle (LKP) whose
nature is model dependent. Candidate LKPs include the first
Kaluza-Klein (KK) excitations of the $U(1)$ gauge boson and the
neutrino ($B^{(1)}$ and $\nu^{(1)}$ respectively). Precision
electroweak measurements \cite{Flacke:2005hb}, the LKP relic
abundance \cite{uedrelicabundance}, and direct-detection
experiments \cite{Servant:2002hb} strongly constrain the viable
ranges of masses for LKPs. However, the allowed range of masses
for the LKP sensitively depends upon the details of the spectrum
of the first and second KK excitations, which can include
significant coannihilation and resonant-annihilation effects. We
compute here the scattering cross section of $\nu^{(1)}$ and of
$B^{(1)}$ off leptons, for which the relevant Feynman diagrams
are shown in Fig.~\ref{fig:feyn}. In the case of the
$B^{(1)}$, we expect large scattering cross sections, since the
intrinsically degenerate nature of the KK spectrum, where
$m_{B^{(1)}}\simeq m_{L^{(1)}}$, clearly enforces a resonant
enhancement. We find, to leading order in $E_l/m_{X}$, and in
the relativistic limit for $l$ and non-relativistic limit for
the LKP particle, that
\begin{eqnarray}
     \sigma_{\nu^{(1)}l}&\simeq
     &\frac{\left|g_{\sss\nu^{(1)}\nu^{(1)}Z}\right|^2}{4\pi
     m_Z^4}\left(g_L^2+g_R^2\right)\ E_l^2,\\ 
     \sigma_{B^{(1)}l}&\simeq &
     \frac{E^2_l}{2\pi}\sum_{R,L}\frac{\left(g_1
     Y_{R,L}\right)^4}{\left(m_{B^{(1)}}^2-m_{l^{(1)}_{R,L}}^2\right)^2},
\end{eqnarray}
where $g_{R,L}$ stand for the $L$ and $R$ couplings of the
lepton $l$ to the $Z^0$ gauge boson, $Y_{R,L}$ for the hypercharge
quantum number, and
$g_{\sss\nu^{(1)}\nu^{(1)}Z}=e/(\sin2\theta_W)$.  The
$\sigma_{Xl}\propto E^2_l$ scaling
found in the case of neutralino dark matter is valid for this
alternative class of WIMPs as well. In the case of the KK
neutrino, further, $\sigma_{\nu^{(1)}l}$ does not depend on the
LKP mass. We stress that consistency with direct-detection
experiments requires $m_{\nu^{(1)}}\gtrsim 50$ TeV
\cite{Servant:2002hb}. While this latter range is in conflict
with estimates of the thermal relic abundance of $\nu^{(1)}$
\cite{uedrelicabundance}, the particle properties of the latter,
assuming the coupling $g_{\sss\nu^{(1)}\nu^{(1)}Z}$ with the $Z^0$,
to be a free parameter instead of
being fixed by the standard gauge interactions, apply to other
dark-matter candidates including the Dirac right-handed neutrino
of 5D warped grand unification \cite{Hooper:2005fj}.

\begin{figure}
\centerline{\epsfig{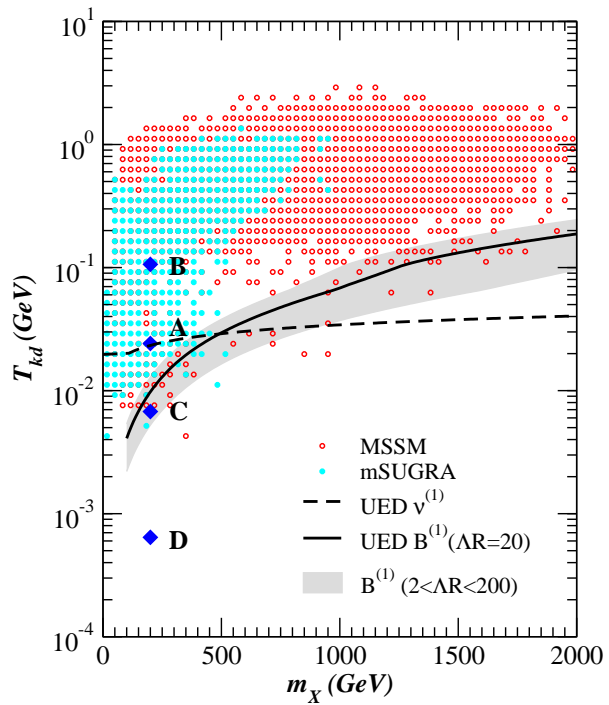}}
\caption{The kinetic-decoupling temperature $T_{\rm kd}$  as a function of the WIMP mass for
     supersymmetric models (red empty dots are for the general MSSM while light-blue filled dots are for mSUGRA) giving a neutralino
     thermal relic abundance consistent with cosmology, and for UED models featuring a
     $B^{(1)}$ and a $\nu^{(1)}$ LKP.  The four benchmark models {\bf A}-{\bf D} discussed in the text are also shown. }
\label{fig:mssm_ued_Tkd}
\end{figure}
\begin{figure}
\centerline{\epsfig{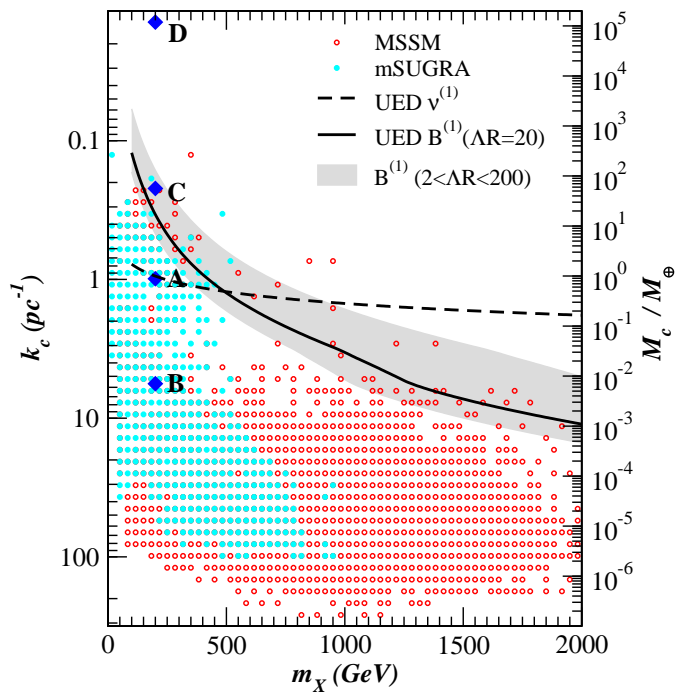}}
\caption{The WIMP protohalo characteristic comoving wavenumber
     $k_c$ (left axis) and mass $M_c$ (right axis) as a function
     of the WIMP mass, for the same models as in
     Fig.~\protect\ref{fig:mssm_ued_Tkd}.} 
\label{fig:mssm_ued_kcut}
\end{figure}

Different WIMP models give rise to different $T_{\rm kd}$, and
therefore to different $M_c$. We apply our elastic-scattering
cross section for WIMPs from light
leptons to the MSSM parameter space, following
the scan procedure of Ref.~\cite{Profumo:2004at},
requiring that the neutralino density falls within the WMAP
2$\sigma$ range for the
dark-matter density \cite{Spergel:2003cb}.
Figs.~\ref{fig:mssm_ued_Tkd} and
\ref{fig:mssm_ued_kcut} show our results for $T_{\rm kd}$
and for $M_c$, respectively, versus the WIMP mass. We also scan
over a subset of the MSSM,
defined by the mSUGRA setup: the results of this second scan are
shown as light blue filled dots in the figures. Finally, we
indicate the range of results expected for a $B^{(1)}$ and a
$\nu^{(1)}$ LKP. We set the KK spectrum according to the minimal
UED prescription for radiative corrections to the KK
masses \cite{radcorr}, setting the cutoff scale $2<\Lambda
R<200$, and showing the ``standard'' $\Lambda R=20$ case with a
black solid line. We also indicate the four
benchmark models of Fig.~\ref{fig:sigmachie} discussed above. We
conclude that (1) in most SUSY models,
$10\lesssim T_{\rm kd}/{\rm MeV}\lesssim4000$, but even lower
values can be attained with finely-tuned resonant channels; (2)
the expected range of $M_c$ varies over the wide
range $10^{-6}\lesssim M_c/M_\oplus \lesssim 10^2$; (3)
$B^{(1)}$ LKPs typically decouple later than neutralinos (around
10 MeV for values of $m_{B^{(1)}}$ preferred by the thermal
relic abundance and by electroweak measurements); and (4) $X$
particles scattering off light leptons through a $Z^0$ exchange
with a coupling equal to $g_{\sss XXZ}$ produce protohalos with
mass
\begin{equation}
     M_{c}\approx 100\ M_\oplus \left|g_{\sss
     XXZ}\right|^{3/2}\left(m_X/100\ {\rm
     GeV}\right)^{-3/4}.
\end{equation}

Since the relevant quantities for WIMP-nucleon scattering and
for the annihilation of WIMPs into gamma rays or antimatter are
typically poorly correlated with WIMP-lepton scattering,
conventional dark-matter-detection rates cannot be
simply related to $T_{\rm kd}$.  However,
we point out that low $T_{\rm kd}$ imply
large WIMP-lepton scattering cross sections and may produce
sizable signals at future electron accelerators using the
search technique recently proposed in
Ref.~\cite{Hisano:2005pm}. If the masses of the sleptons and
neutralino can be measured by future colliders then, if $\Delta
m$ is small enough, the beam energy of a future electron
accelerator might be tuned to $E_{\rm beam}\simeq\Delta m$,
enabling resonant $s$-channel scattering of neutralino
dark-matter. The scattered electrons could then be detected by
calorimeters or tracking chambers along the beam line. This
technique would also reveal information about the dark-matter
velocity distribution \cite{Hisano:2005pm}. As a rule of thumb,
more than one event per year is expected at a future electron
collider with 100 m of detector length and 10 A of beam current
if $T_{\rm kd}\lesssim 10$ MeV; in the extreme case of $T_{\rm
kd}\approx 1$ MeV, the expected event rate per year could be as
large as  $\sim 10^4$!

The temperature $T_{\rm kd}$ in WIMP models has a critical
impact not only on the size distribution of primordial
protohalos expected in $N$-body simulations of structure
formation \cite{Diemand:2005vz}, but also for potential effects
in WIMP direct and indirect detection induced by dark matter
clumps or streams, or in the anisotropy of the cosmic gamma-ray
background induced by WIMP annihilations
\cite{Ando:2005xg}. Accounting for the wide range of
possibilities consistent with detailed WIMP models
should therefore be regarded as an essential ingredient for
future studies in this field. 

\begin{acknowledgments}

KS is supported by NASA through Hubble Fellowship grant
HST-HF-01191.01-A awarded by the Space Telescope Science
Institute, which is operated by the Association of Universities
for Research in Astronomy, Inc., for NASA, under contract NAS
5-26555.  SP and MK are supported in part by DoE
DE-FG03-92-ER40701 and FG02-05ER41361 and NASA NNG05GF69G.

\end{acknowledgments}


\begin{thebibliography}{100}

\bibitem{WIMPReviews}
  G.~Jungman, M.~Kamionkowski, and K.~Griest,
  Phys.\ Rept.\  {\bf 267}, 195 (1996) [arXiv:hep-ph/9506380];
  L.~Bergstrom, Rept.\ Prog.\ Phys.\ {\bf 63}, 793 (2000)
  [arXiv:hep-ph/0002126];
  G.~Bertone, D.~Hooper, and J.~Silk,
  Phys.\ Rept.\  {\bf 405}, 279 (2005) [arXiv:hep-ph/0404175].

\bibitem{Boehm} C. B\oe hm, P. Fayet, and R. Schaeffer,
      Phys. Lett. B {\bf 518}, 8 (2001) [arXiv:astro-ph/0012504].

\bibitem{Chen:2001jz}
  X.~L.~Chen, M.~Kamionkowski, and X.~M.~Zhang,
  Phys.\ Rev.\ D {\bf 64}, 021302 (2001) [arXiv:astro-ph/0103452].

\bibitem{GreenWIMPy}
  A.~M.~Green, S.~Hofmann, and D.~J.~Schwarz,
  Mon.\ Not.\ Roy.\ Astron.\ Soc.\  {\bf 353}, L23 (2004)
  [arXiv:astro-ph/0309621];
  A.~M.~Green, S.~Hofmann, and D.~J.~Schwarz,
  JCAP {\bf 0508}, 003 (2005) [arXiv:astro-ph/0503387].

\bibitem{Diemand:2005vz}
  J.~Diemand, B.~Moore, and J.~Stadel,
  Nature {\bf 433}, 389 (2005) [arXiv:astro-ph/0501589];
  J.~Diemand, M.~Kuhlen and P.~Madau,
  arXiv:astro-ph/0603250.

\bibitem{Ando:2005xg}
  S.~Ando and E.~Komatsu,
  Phys.\ Rev.\ D {\bf 73}, 023521 (2006)
  [arXiv:astro-ph/0512217].

\bibitem{Loeb:2005pm}
  A.~Loeb and M.~Zaldarriaga,
  Phys.\ Rev.\ D {\bf 71}, 103520 (2005) [arXiv:astro-ph/0504112].

\bibitem{msugra}
  A.~H.~Chamseddine, R.~Arnowitt, and P.~Nath,
  Phys.\ Rev.\ Lett.\  {\bf 49}, 970 (1982).

\bibitem{nonuniv}
  H.~Baer {\it et al.},
  JHEP {\bf 0406}, 044 (2004)
  [arXiv:hep-ph/0403214];   S.~Profumo,
  Phys.\ Rev.\ D {\bf 68}, 015006 (2003)
  [arXiv:hep-ph/0304071].

\bibitem{ued}
  T.~Appelquist, H.~C.~Cheng, and B.~A.~Dobrescu,
  Phys.\ Rev.\ D {\bf 64}, 035002 (2001)
  [arXiv:hep-ph/0012100];  G.~Servant and T.~M.~P.~Tait,
  Nucl.\ Phys.\ B {\bf 650}, 391 (2003)
  [arXiv:hep-ph/0206071].

\bibitem{Flacke:2005hb}
  T.~Flacke, D.~Hooper, and J.~March-Russell,
  arXiv:hep-ph/0509352.

\bibitem{uedrelicabundance}
  K.~Kong and K.~T.~Matchev,
  JHEP {\bf 0601}, 038 (2006)
  [arXiv:hep-ph/0509119];   F.~Burnell and G.~D.~Kribs,
  Phys.\ Rev.\ D {\bf 73}, 015001 (2006)
  [arXiv:hep-ph/0509118];   M.~Kakizaki {\it et al.},
  Nucl.\ Phys.\ B {\bf 735}, 84 (2006)
  [arXiv:hep-ph/0508283].

\bibitem{Servant:2002hb}
  G.~Servant and T.~M.~P.~Tait,
  New J.\ Phys.\  {\bf 4}, 99 (2002)
  [arXiv:hep-ph/0209262].

\bibitem{Hooper:2005fj}
  K.~Agashe and G.~Servant,
  JCAP {\bf 0502}, 002 (2005)
  [arXiv:hep-ph/0411254]; D.~Hooper and G.~Servant,
  Astropart.\ Phys.\  {\bf 24}, 231 (2005)
  [arXiv:hep-ph/0502247].

\bibitem{Profumo:2004at}
  S.~Profumo and C.~E.~Yaguna,
  Phys.\ Rev.\ D {\bf 70}, 095004 (2004)
  [arXiv:hep-ph/0407036].

\bibitem{Spergel:2003cb}
  D.~N.~Spergel {\it et al.}  (WMAP Collaboration),
  Astrophys.\ J.\ Suppl.\  {\bf 148}, 175 (2003)
  [arXiv:astro-ph/0302209].

\bibitem{radcorr}
  H.~C.~Cheng, K.~T.~Matchev, and M.~Schmaltz,
  Phys.\ Rev.\ D {\bf 66}, 036005 (2002)
  [arXiv:hep-ph/0204342].

\bibitem{Hisano:2005pm}
  J.~Hisano {\it et al.},
  AIP Conf.\ Proc.\  {\bf 805}, 423 (2006)
  [arXiv:hep-ph/0504068].

\end{thebibliography}
\end{document}